\documentclass[useAMS,usenatbib,usegraphicx]{mn2e}

\title[Removing residual OH emission from near infrared spectra]{A method to remove residual OH emission from near infrared spectra\thanks{Makes use of
    observations at the European Southern Observatory VLT (074.A-9011 and 075.B-0040).}}

\author[R. I. Davies]{R. I. Davies\thanks{E-mail: davies@mpe.mpg.de}\\
Max Planck Institut f\"ur extraterrestrische Physik, Postfach 1312, Garching, 85741, Germany}
\begin{document}


\pagerange{\pageref{firstpage}--\pageref{lastpage}} 

\maketitle

\label{firstpage}

\begin{abstract}
I present a technique to remove the residual OH airglow emission from
near infrared spectra.
Historically, the need to subtract out the strong and variable OH
airglow emission lines from 1--2.5\,micron\ spectra has imposed severe
restrictions on observational strategy.
For integral field spectroscopy, where the field of view is limited,
the standard technique is to observe blank sky frames at regular
intervals.
However, even this does not usually provide sufficient compensation if
individual exposure times are longer than 2--3\,minutes due to 
(1) changes in the absolute flux of the OH lines,
(2) variations in flux among the individual OH lines, and 
(3) effects of instrumental flexure which can lead to `P-Cygni' type
residuals.
The data processing method presented here takes all of these effects
into account and serendipitously also improves background subtraction
between the OH lines.
It allows one, in principle, to use sky frames taken hours or days
previously so that observations can be performed in a quasi-stare mode.
As a result, the observing efficiency (i.e. fraction of time spent on
a source) at the telescope can be dramatically increased.
\end{abstract}


\section{Introduction}
\label{sec:intro}

Near infrared airglow emission originates in OH radicals which are
created by reactions between ozone and hydrogen high in the atmosphere.
Removing the emission lines which result from the subsequent radiative
cascade is a crucial part of processing near infrared (1--2.5\micron)
spectra.
While they are usually considered to be a nuisance in this respect,
they also provide a useful reference for wavelength calibration
\citep{oli92,mai93,ost96,rou00}.

\cite{mai93} has measured the strongest lines, which lie in the
H-band, to have fluxes of order
400\,ph\,s$^{-1}$\,m$^{-2}$\,arcsec$^{-2}$.
This contrasts strongly with the background continuum they measured
between the lines of only
590\,ph\,s$^{-1}$\,m$^{-2}$\,arcsec$^{-2}$\,\micron$^{-1}$.
These values demonstrate that, even at a moderate spectral resolution of
$R\sim3000$, the background level on an OH line can be more than 3
orders of magnitude higher than that between them.
There are inevitably implications both for the statistical photon noise 
and also for systematic effects when attempting to remove the OH emission by
subtracting a `sky' spectrum from an `object' spectrum.
The first of these noise sources is unavoidable, and can only be
improved by integrating longer.
The latter is an issue because the OH line fluxes can vary
significantly on timescales of only a few minutes;
and yet for a clean subtraction, the flux of the OH lines in
the sky and object frames would have to differ by much less than 1\%.

The flux variations are generally not a problem for longslit spectra,
because the slit 
is usually much more extended than the object of interest: 
120\arcsec\ in the case of ISAAC \citep{moo98} at the VLT.
With such data, the OH emission can be completely removed from the
rectified 2-dimensional spectrum:
at each spectral pixel, one
can fit a function to the residual background along the spatial
dimension and subtract it.

Near infrared integral field spectrometers do not allow this because
of the restricted field of view: there is no point in the observed
field of view more than a few arcsec from the centre.
For example, the largest field of view of SINFONI
\citep{eis03a,eis03b,bon04} at the VLT is $8\arcsec\times8\arcsec$;
and at the finest pixel scale used with adaptive optics it is less than
1\arcsec.
Similarly, OSIRIS \citep{lar06,kra06} for use with adaptive optics at
the Keck II telescope has a field 
of view ranging from $0.32\arcsec\times1.28\arcsec$ to a maximum of
$4.8\arcsec\times6.4\arcsec$ depending on pixel scale and spectral
coverage.
And the GNIRS IFU \citep{all06} for Gemini South telescope has a field
of view of $3.15\arcsec\times4.46\arcsec$.
The KMOS \citep{sha05} multiple integral field unit which is being
designed for the VLT will have individual fields of
$2.8\arcsec\times2.8\arcsec$. 
It is clear that in general astronomical targets will fill a substantial
fraction -- and perhaps all -- of the field in any of these IFUs.
As a result, one cannot easily extrapolate the
background level from pixels at the edge of the field of view to those
in the centre.
For example, in crowded stellar fields such as the Galactic Centre,
the combination of nebular emission, dust continuum, and faint stars
make it challenging to identify any region of pure background in the
central few arcsec.
If one observes (active) galactic nuclei, although the emission may be
dominated by a bright compact nucleus, there is still easily detectable
extended emission out to more than a few arcsec.
And in high redshift galaxies which one may expect to be only
$\sim1$\arcsec\ across, there may be line emission on larger scales --
for example \cite{for06} were able to detect line emission across
2--3\arcsec\ in several $z>2$ galaxies.
Thus in most cases it is mandatory to observe separate blank sky fields.

Compounding the problem is the demand to reach the maximum
signal-to-noise even between the OH lines which requires exposure
times of individual frames be 5--10\,mins, to avoid their being
read-noise limited.
Hence, even if one takes frequent sky frames, the OH lines are almost
always imperfectly subtracted.

The situation is complicated further by the fact that in such
instruments the spectral line profile can vary across the field of
view due to image distortions, off-axis aberrations, and manufacturing
variations between elements.
Thus one cannot simply take a sky spectrum from one spatial pixel and
subtract it from another.
This is also the reason why attempts to generate model OH spectra that
can be subtracted from the data inevitably fail.
To make matters worse, at moderate resolution the strength of the OH
lines means that even small amounts of spectral flexure in an instrument
(corresponding to only a few percent of a resolution element) can leave a
significant `P-Cygni' shape residual when a sky frame is subtracted.

The rather complicated topic of background subtraction for integral
field spectrometers has been addressed by \cite{all99}.
To reduce the effects of temporal uncertainty in the background, they
proposed osberving a simultaneous blank sky field.
This paper explores an alternative technique for removing
the variable OH airglow without the need for a simultaneous sky frame.

\section{Compensating for variable OH line fluxes}
\label{sec:ohsub}

\begin{figure*}
\includegraphics[width=10cm]{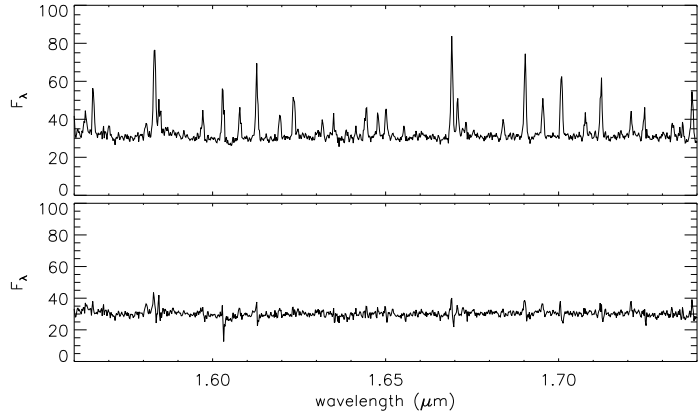}
\caption{Top: Spectrum of a sky subtracted 5\,min SINFONI exposure
  showing residual OH lines after subtracting a sky frame taken
  immediately afterwards.
The flux in these residuals is only $\sim3$\% of the original OH emission, but
  is still significant.
Bottom: the same two frames are subtracted using the scaling procedure
  described in the text in Section~\ref{sec:ohsub}.
The spectra are of NGC\,3783 and in both cases were extracted within a
  0.5\arcsec\ aperture centered slightly off the galaxy nucleus.
The reduction in noise is discussed quantitatively in
  Section~\ref{sec:stare} and shown in Fig.~\ref{fig:qstareh}.}
\label{fig:ohsub}
\end{figure*}

In this section a method which allows one to compensate for variations
in both absolute and relative flux of the OH lines between object and
sky frames is presented.
These variations give rise to residuals such as those in
Fig.~\ref{fig:ohsub} which, although only a few
percent the original line strength, are still very significant.

The basis of the technique is to find a
scaling as a function of wavelength that can be applied to a spectrum
extracted from a sky cube in order to match it optimally to the sky
background in an object cube.
The scaling function is then applied separately to the spectrum at
each spatial position of the sky cube individually, creating a
modified sky cube.
It is the entire modified sky cube that is then subtracted from the
object cube -- conveniently avoiding the issue of the variable
spectral line profiles.

The scaling function can be found relatively easily if one understands
the origin of the variation in the OH line fluxes.
The relative populations of the excited states in the OH radical can
be approximated well by Boltzmann distributions characterised by
rotational and vibrational temperatures (e.g. \citealt{rou97}).
A rotational temperature of 205$\pm$5\,K was found by \cite{wil96}
in Antarctica, which corresponds well to models of the kinetic
temperature at the 85\,km altitude where OH radicals exist.
Although the vibrational temperature would have to be unphysically
high, in the range 8500--13000\,K \citep{wal68}, it does provide a good
representation of the intensities in the different bands.

\subsection{Vibrational Variations}
\label{sec:vibvar}

\begin{figure*}
\includegraphics[width=14cm]{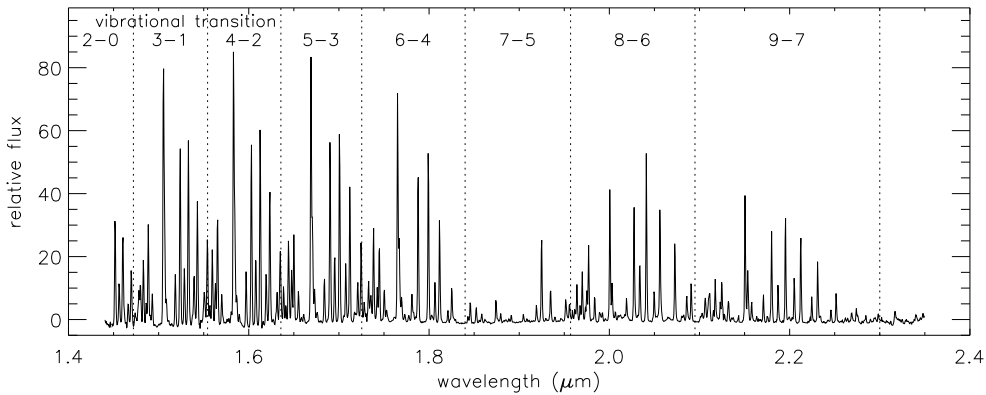}
\caption{Spectrum of the OH emission across the H- and K-bands (the
  thermal background has been subtracted).
The various spectral ranges corresponding to the different vibrational
  transitions (labelled) are indicated.
Very few OH lines from the $7-5$ transition are observed due to
  atmospheric absorption (between the H- and K-bands) at these
  wavelengths.}
\label{fig:segments}
\end{figure*}

It is between the vibrational bands that most of the variation in the
OH spectrum occurs.
Fortuitously, transitions between any two particular vibrational bands
tend to lie, to a large extent, within well defined wavelength limits.
These are shown for the H-and K-bands in Fig.~\ref{fig:segments}.
Furthermore, there is only some small overlap between different
vibrational transitions; and within the overlap, the OH lines
originating from either one or the other vibrational
transition tend to be relatively weak.
Thus, to a first approximation one can divide the spectrum into sections
corresponding to specific vibrational transitions, and treat these
separately.

\begin{figure*}
\includegraphics[width=12cm]{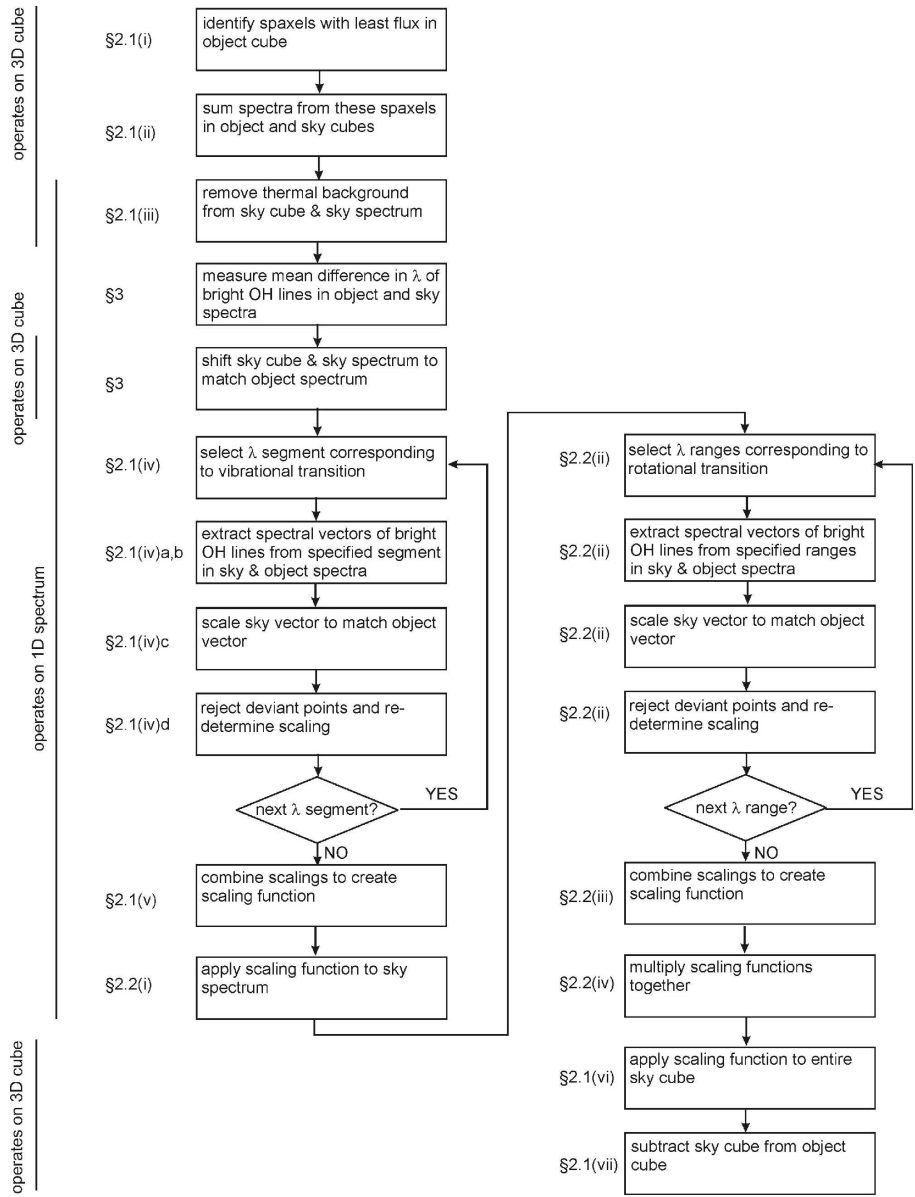}
\caption{Flowchart showing the steps in the combined algorithm to
  correct both for changes in the vibrational and rotational
  temperature of the OH radical, and also for spectral flexure.
 Next to each step is indicated the place from where it comes in the
  text.
Also shown is whether each step applies to a summed 1-dimensional
  spectrum or to the entire data cube.}
\label{fig:flowchart}
\end{figure*}

It is this that the algorithm outlined below has been developed to do.
A flow chart which includes these steps is given in Fig.~\ref{fig:flowchart}.

\begin{enumerate}

\item
identify the spatial pixels in the object cube which contain the least
flux (typically 50\% of the spatial pixels are selected).

\item
sum spectra from these positions in both the object and sky cubes to
create an object spectrum and a sky spectrum.

\item
Fit a blackbody function to the thermal background in the sky
spectrum, and subtract it from the sky spectrum as well as the
original sky cube (to leave just the line emission).

\item
for each spectral segment (corresponding to each vibrational
transition) do the following:

\begin{enumerate}
\item
extract a vector array containing the few spectral pixels around each
of the major OH lines in the sky 
segment and concatenate them into a single array (equivalently one can
select all pixels which have a flux above some fraction, such as 20\%,
of the maximum in that segment).

\item
Repeat the extraction and concatenation for the object segment using
the same pixels, but also subtracting the continuum 
for each OH line (it can be interpolated from spectral regions immediately
either side of each line).

\item
find the scaling factor which minimizes the difference between the sky and
object vectors.

\item
if there are any pixels in the object vector which deviate
significantly from zero when the scaled sky vector is subtracted,
reject them and repeat steps (a)--(c) above. This makes the procedure
robust against any line emission from the science target that
coincides with an OH line.

\end{enumerate}

\item
combine the scalings from each segment to generate the vibrational scaling
function across the whole spectrum.

\item
multiply the spectrum at each spatial position in the sky cube (from
which the thermal background has been subtracted) by the
scaling function to create a modified sky cube.

\item
subtract the modified sky cube from the object cube.

\end{enumerate}

The procedure as given above is not specific to OH lines, and hence
can also be used for the jungle of emission features longward of
2.35\micron\ in the K-band as well as the O$_2$ emission in the J-band.

An additional advantage comes from the fact that the scaling for each
segment is derived from the stronger emission lines but
is applied to the whole segment.
This means that the weak unresolved OH continuum is also
scaled similarly, providing a better quality background subtraction
even between the obvious emission lines.

\subsection{Rotational Variations}
\label{sec:rotvar}

\begin{figure*}
\includegraphics[width=12cm]{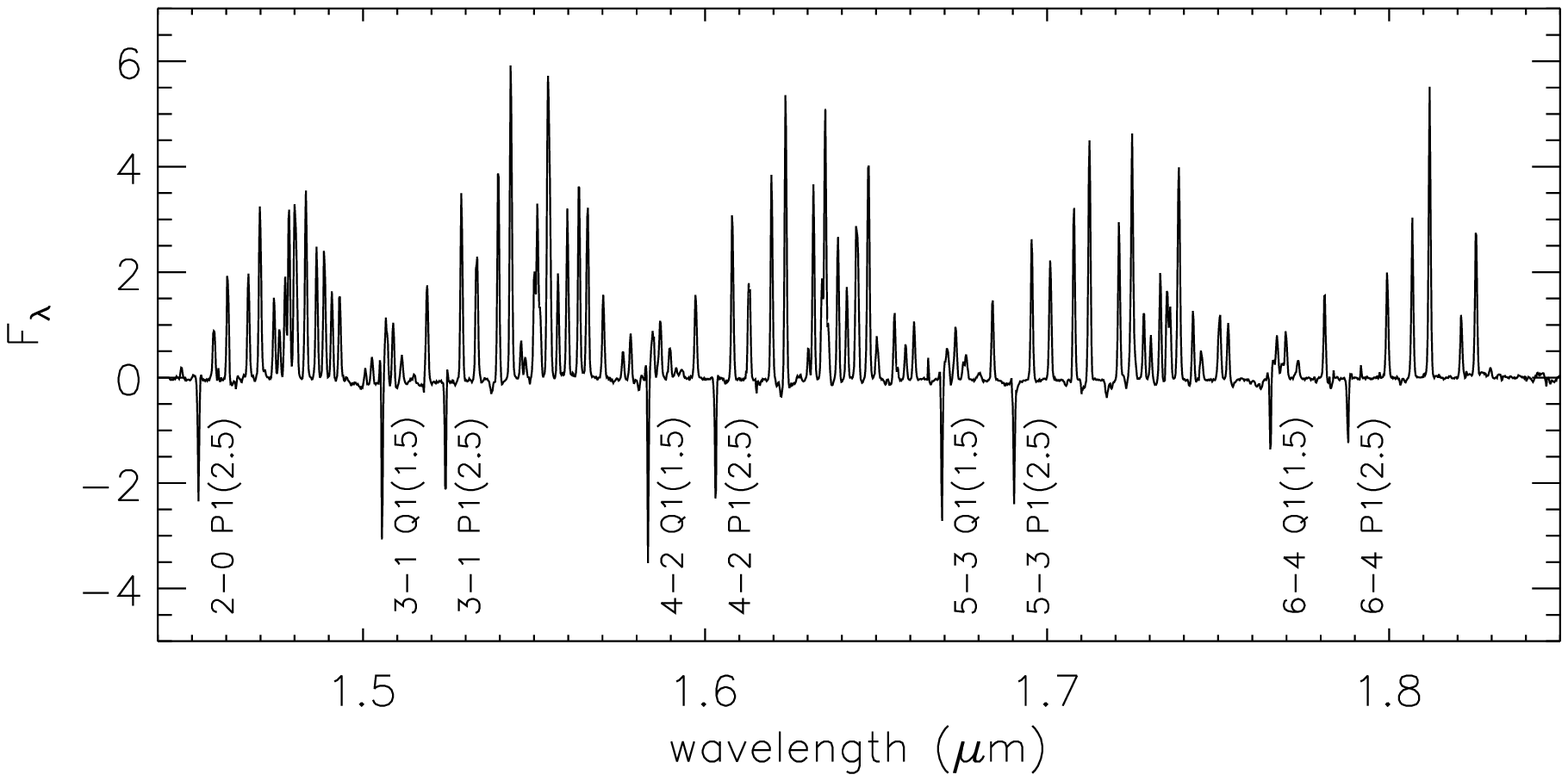}
\caption{Spectrum of a 5\,min SINFONI exposure which has had the sky
  subtracted using the algorithm described in Section~\ref{sec:ohsub}
  to account for changes in the vibrational temperature.
The residual shown is typical of what is seen with a change in the
  rotational temperature: the lines from the lowest upper level --
  Q1(1.5) and P1(2.5) -- are over- (or equivalently under-) subtracted.}
\label{fig:rotres}
\end{figure*}

Refining the procedure to take into account changes in the rotational
temperature of the OH radicals can quickly become rather more complex
at the spectral resolutions of $R\sim3000$ considered here.
The reason is that there are many emission lines from different
rotational levels whose wavelengths coincide and are blended at this
resolution.
Nevertheless, there are still a few distinct groups of lines which can
be scaled together.
Fig.~\ref{fig:rotres} shows a residual -- after correcting the sky
subtraction for changes in the vibrational temperature as already
described in Section~\ref{sec:vibvar} -- typical of a change in the
rotational temperature.
The lines from the lowest upper rotational level of the 
$X^2\Pi_{3/2}$ sub-state, the P1(2.5) and Q1(1.5) transitions from the 
$J_{up} = 3/2$ level (see Fig.~2 of \citealt{rou00}), are over-subtracted.
This means that this group of lines can be corrected together -- hence
preserving robustness against line emission from the object occuring at the
same wavelength as an OH line -- using a
single additional scaling which can be computed in a way analagous to
that described previously.

If this is done, one then sees a new set of lines over-subtracted:
the P1(3.5), Q1(2.5), and R1(1.5) rotational transitions from the 
$J_{up} = 5/2$ level.
This is where the difficulties begin, because for each vibrational
transition the Q1(2.5) lines are
only 10--20\AA\ from the Q1(1.5) lines.
At a spectral resolution of $R\sim3000$ these are only a few
resolution elements apart and hence still blended in their wings.
Thus one cannot cleanly isolate them in order to apply an appropriate
correction.
Nevertheless, one can still make a significant improvement by simply
excluding the Q1(2.5) lines from this group and correcting the
others.

Thus a straight-forward implementation can be made using three groups of
rotational transitions:
the P1(2.5) and Q1(1.5) lines; 
the P1(3.5) and R1(1.5) lines;
and the rest of the spectrum (the scaling for this being determined
from all the remaining isolated lines).
This leads to three scalings which can be multiplied into their respective
parts of the vibrational scaling function to yield the final scaling
function.
The algorithm, which is included in the flow
chart shown in Fig.~\ref{fig:flowchart}, is then as follows:

\begin{enumerate}

\item
having created a vibrational scaling function, multiply this into the
sky spectrum.

\item
for each set of wavelength ranges corresponding to the rotational
transitions given above, perform exactly the same steps as described
in (iv) a--d of Section ~\ref{sec:vibvar}.

\item
combine the scalings from each set of wavelength ranges to generate
the rotational scaling function across the whole spectrum.

\item
multiply the vibrational and rotational scaling functions together.

\item
continue at step (vi) of Section ~\ref{sec:vibvar}, multiplying the
whole sky cube by the combined scaling function

\end{enumerate}

While this does not provide a full correction for changes in the
rotational temperature, it can -- particularly in the H-band -- make
a significant enhancement on correcting only for the vibrational
temperature.

It is worth noting that a simpler alternative becomes possible if one
has simultaneous
observations of a blank sky region (see also \citealt{all99}), as is
possible with multi-object spectrometers.
One could then target each line -- or blend of lines -- independently,
irrespective of its transition, using a similar scheme to that
described above:
derive the respective scalings from the simultaneous sky observation
(in a different slitlet or IFU) and apply them to the sky frame taken
previously (in the same slitlet or IFU as the object frame).

\section{Compensating for Instrumental Flexure}
\label{sec:flexure}

\begin{figure*}
\includegraphics[width=10cm]{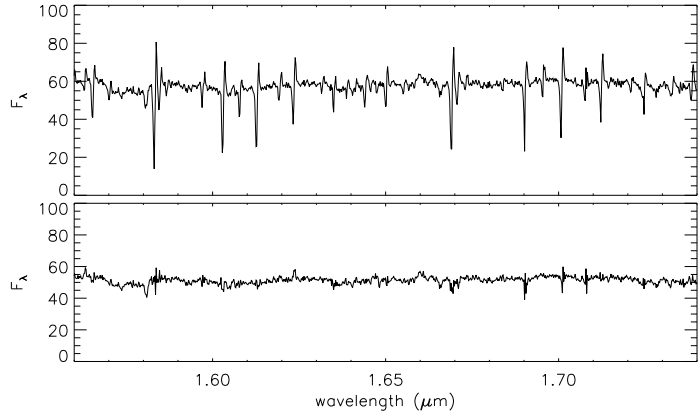}
\caption{Top: Spectrum of a sky subtracted 5\,min SINFONI exposure showing
  P-Cygni OH residuals, characteristic of spectral flexure between the
  object and sky frames.
Bottom: the same two frames are subtracted using the flexure
  compensation procedure described in the text in
  Section~\ref{sec:flexure}.
The spectra are of NGC\,3783 and in both cases were extracted within a
  0.5\arcsec\ aperture centered slightly off the galaxy nucleus.
The reduction in noise is discussed quantitatively in
  Section~\ref{sec:stare} and shown in Fig.~\ref{fig:qstareh}.}
\label{fig:pcygni}
\end{figure*}

\begin{figure*}
\includegraphics[width=10cm]{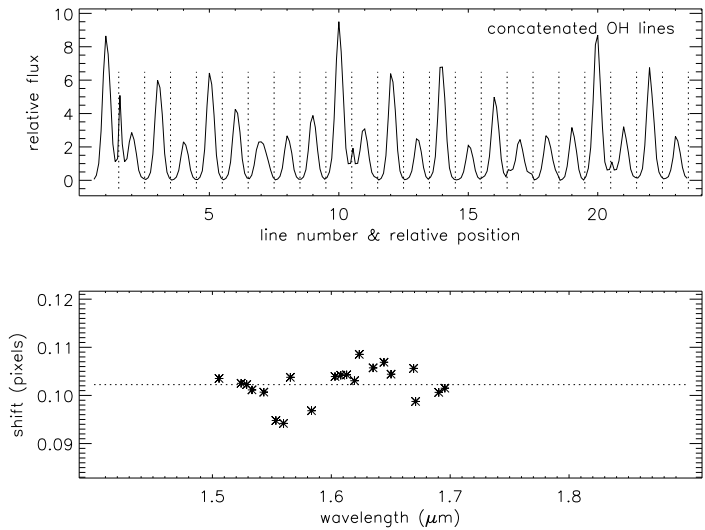}
\caption{Top: concatenated vector of spectral segments containing
  strong OH lines which are used to
  determine the spectral shift between the object and sky frames.
The shift between sky and object frames for each OH line in this
  array corresponds to one data point in the bottom panel.
Bottom: Measured offset as a function of wavelength, each point
  corresponding to one OH line in the vector above.
This method allows one to measure the mean offset with an accuracy of at
  least 0.01\,pixel; and also to detect any trend with wavelength that
  there may be.}
\label{fig:offset}
\end{figure*}

Very often, particularly in instruments mounted at a Cassegrain focus
or those at a Nasmyth focus which need to rotate, there can be flexure
resulting in shifts of the wavelength scale between exposures.
Here we are concerned only with spectral shifts (regardless of the
actual cause), since even small shifts can have a dramatic impact on the
quality of the OH line subtraction.
For example, subtracting 2 lines with the same flux but shifted by
only 0.1
times the FWHM can leave a P-Cygni shape residual with an amplitude of
more than 10\% of the original line.
A case where this has occured in SINFONI data is shown in
Fig.~\ref{fig:pcygni}.
Once such residuals are created -- for example by subtracting a sky
frame from an object frame without correcting the shift {\em a priori}
-- they are extremely difficult to correct.

The obvious remedy is to apply a shift to the sky frame before
subtracting it from the object frame.
However, raw frames are not appropriate for this operation for 2 reasons:
firstly, there will be numerous bad pixels which would normally be
subtracted out but instead will propagate during
the necessary interpolation;
secondly, it is difficult to measure the required shift from raw
spectra which contain significant curvature.
Instead, a sequence of alternative data processing steps is outlined
below:

\begin{enumerate}

\item
subtract dark frames from both object and sky
frames (rather than subtracting sky frames from the object frames
directly) so that the OH emission remains in every frame.

\item
reconstruct the object and sky cubes from the dark subtracted frames.

\item
create a `thermal background' cube from each sky cube (e.g. by running
the routine described in Section~\ref{sec:ohsub} but using each sky
frame as both the `object' and `sky' input).

\item
subtract the appropriate thermal background cubes from each object
cube (if the thermal background cubes do not change much, it may be
possible to average them all to improve the signal-to-noise).

\item
remove the OH emission from each object cube by using the routine
described in Section~\ref{sec:ohsub}, but also applying a shift to the
wavelength scale of the respective sky cube as described below.

\end{enumerate}

A method which has proven successful for measuring spectral shifts is
to determine the position of every strong OH line in both the sky and
object spectra -- either by calculating the line centroids or by
fitting Gaussian functions.
This then yields a series of offsets as a function of wavelength, as
shown in Fig.~\ref{fig:offset}.
It is usually sufficient to take the iterative mean of these,
rejecting outliers which deviate significantly, and apply it as a
constant shift to the entire sky spectrum.
As indicated in Fig.~\ref{fig:flowchart}, a couple of steps to measure
and apply this shift can easily be included immediately 
after step (iii) of the routine described in Section~\ref{sec:vibvar}.
The same shift would also have to be applied to the complete modified
sky cube before subtracting it in step (vii) of the same routine.


\section{Towards Efficient Observing: a Quasi-stare mode}
\label{sec:stare}

\begin{figure*}
\includegraphics[width=12cm]{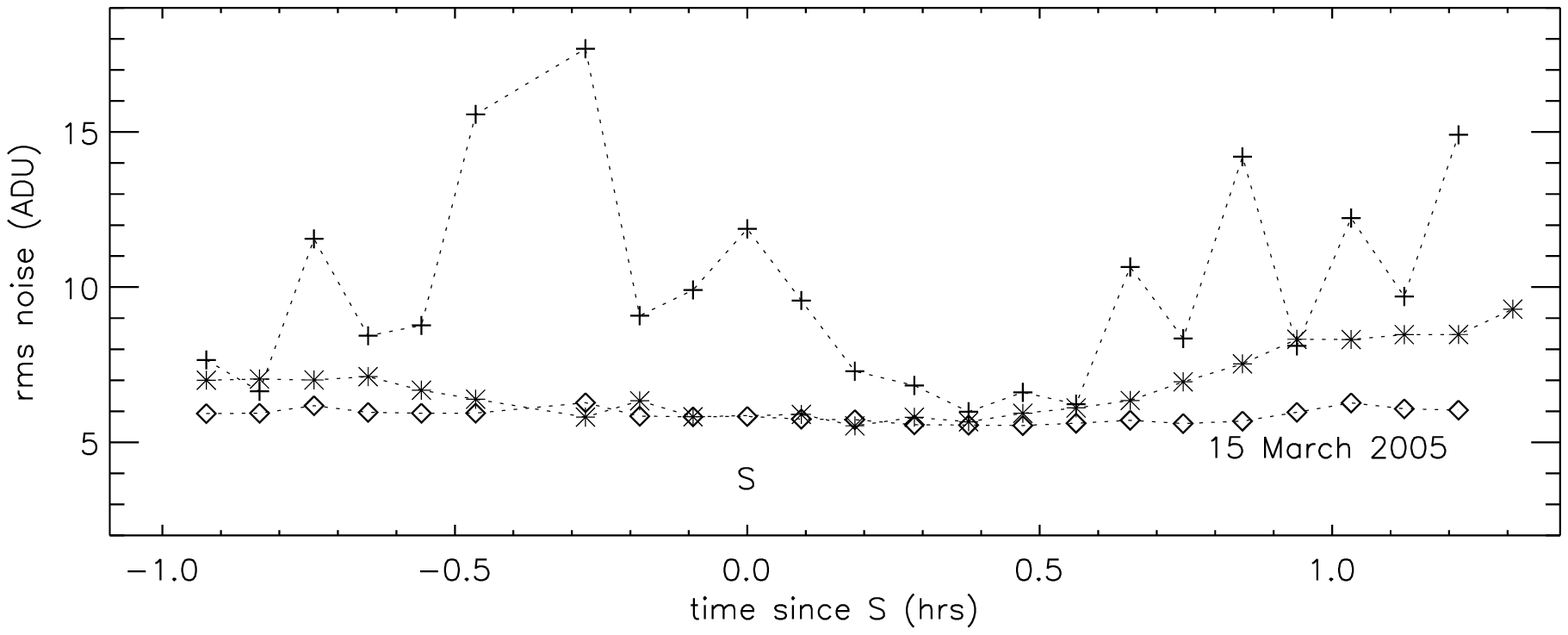}
\caption{Noise in blank sky subtracted H-band SINFONI cubes (arbitrary
  data units) for consecutive 5\,min integrations.
The noise is measured across the whole field of view and in the
  wavelength range 1.55-1.75\micron.
Pluses denote cubes which each had a different sky cube, taken
  immediately afterwards, subtracted.
Diamonds are the same, except the sky subtraction was performed using
  the routines described in this paper.
Asterisks denote cubes from which the same sky cube (taken at the
  time indicated by `S') was subtracted, also using these routines.
The increase in noise for frames observed more than
30\,mins after S is due to
variations in the rotational temperature of the OH lines, which is
only partially compensated.}
\label{fig:qstareh}
\end{figure*}

\begin{figure*}
\includegraphics[width=12cm]{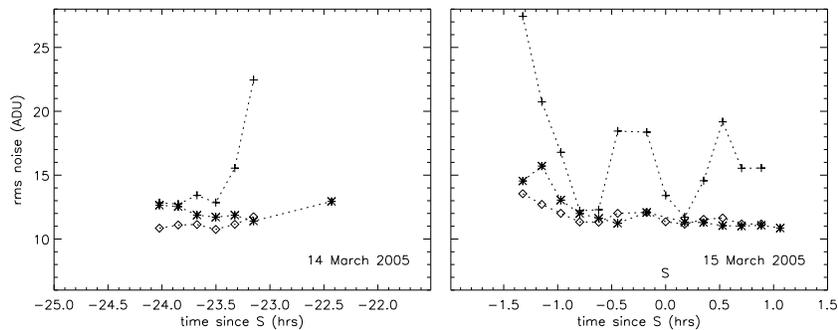}
\caption{Noise in blank sky subtracted K-band SINFONI cubes (arbitrary
  data units) 
  for consecutive 10\,min integrations taken on 2 nights.
The noise is measured across the whole field of view and in the
  wavelength range 2.00--2.25\micron\ (where the OH emission is most
  prominent).
Pluses denote cubes which each had a different sky cube, taken
  immediately afterwards, subtracted.
Diamonds are the same, except the sky subtraction was performed using
  the routines described in this paper.
Asterisks denote cubes from which the same sky cube (taken at the
  time indicated by `S') was subtracted, also using these routines.}
\label{fig:qstarek}
\end{figure*}

In this section the performance of the strategies above are
illustrated quantitatively using a sequence of 24 H-band SINFONI
frames, each a 5\,min exposure, taken during October 2005.
These are nodded observations of a $z=1.4$ galaxy which is spatially
compact (occupying less than 1\arcsec\ of the 8\arcsec\ field width)
and has a rather faint continuum so that in a single exposure it is
barely detected.
Thus, for the purpose here, these frames in effect form a series of
consecutive blank sky frames.
Fig~\ref{fig:qstareh} shows the noise in the reconstructed cubes.
The sky was subtracted in 3 ways.
In one instance, for each frame the subsequent one was used as the
`sky' and simply subtracted, yielding 23 independent cubes.
The mean noise in these was 9.91\,ADU, but there is a great deal of
variation depending on whether the OH lines happen to subtract well or
not.
For the second instance, the routines described in
Sections~\ref{sec:ohsub} and \ref{sec:flexure} were used to optimise
the subtraction.
The mean noise in each of these cubes was much less, being only
5.86\,ADU, and much more stable; 
when the first 9 were combined, the noise in the resulting cube was
1.99\,ADU, close to that expected.
In the third instance, only a single sky frame (close to the middle of
the sequence, denoted by `S' at time zero) was subtracted from every
other cube using the routines described previously.
In all cases, the noise is less than that reached by simply
subtracting the nearest sky frame, and for a period of about 30\,mins
before and after the sky frame was taken, the noise in the resulting
cubes was comparable to that of the cubes processed optimally with
their own sky frames.
Combining 9 of these frames (with small 2 pixel spatial dithers
between them) yielded a cube with a noise of 2.07\,ADU.

From these results one can conclude that even when using sky
frames taken after each object exposure, the background subtraction
can be improved -- often significantly -- by using the algorithms
described here.
Furthermore, they enable one to achieve comparable or better background
subtraction even when using a sky frame taken at a different time.
Thus, a similar noise level can been reached but in a much shorter
amount of observing time: in the example above, one needs only a
single sky frame rather than 9.
Hence the observing efficiency can be increased from 50\% to 90\% without
loss of signal-to-noise.

The increase in noise for the frames which were observed more than
30\,mins before or after the sky frame is due to
variations in the rotational temperature of the OH lines -- which is
only partially compensated in the scheme described in this paper.
In the H-band, where the OH lines are strongest, this appears to be
a limiting factor.
On the other hand, a similar test with K-band data
have shown that in most cases one
can use sky cubes taken on different nights without losing
signal-to-noise.
Fig.~\ref{fig:qstarek} shows the results of the test involving 21
exposures of 10\,min taken over 2 nights.
Again, these are nodded observations of a high redshift galaxy
at $z=2.4$ which is compact (also filling less than 1\arcsec\ of the
8\arcsec\ field width) and is only detected even in line emission
after multiple integrations -- thus
an individual exposure is effectively blank.
The noise in each of the resulting cubes has been measured in the
wavelength range 2.00--2.25\micron, 
where the OH emission is most prominent and the thermal background is
relatively low.
Individual frames which each had their own separate sky cube simply
subtracted had a mean noise of 16.11\,ADU across both nights, but with
considerable variation as was found in the H-band test.
When the sky subtraction was optimised, the mean noise was reduced to
11.57\,ADU and the stability significantly improved.
When 7 of these optimised cubes are combined, the noise in the
resulting cube was 4.56\,ADU, close to the expected value.
Cubes from which a single sky cube taken on the same night was
subtracted had a mean noise of 12.14\,ADU; even when the sky cube was
observed on a different night, the mean noise in the processed cubes
was still the same at 12.06\,ADU.
In all cases an optimal sky subtraction using the single frame was
better than a simple subtraction using a frame taken immediately
afterwards.
When 7 such cubes from each night are combined (using 2\,pixel spatial
dithers), the resulting noise is 6.09\,ADU and 6.07\,ADU.
Thus by using only a single sky frame one loses slightly in terms of
signal-to-noise but gains significantly in terms of telescope time:
each of the latter combined cubes required only 80\,mins of open
shutter time, whereas the former combined cube required 140\,mins.

Based on the tests above, it seems reasonable to conclude that it is
not necessary to observe a sky frame immediately
before or after each object frame -- i.e. that one does not need to 
spend 50\% of the time observing blank sky, as has
historically been done.
Sky subtraction works equally well using a frame taken 30--60\,mins
before (and for K-band the frames can in principle be taken days apart).
A strategy in
which a sky frame is taken only once per hour rather than every other
exposure would represent a significant step towards
observing in a `stare' mode.
As long as there are small (e.g. 2 pixel) dithers between the
pointings of each object frame, the noise in the sky frame will not add
coherently when all the sky-subtracted cubes are combined.
This is valid as long as one wishes to preserve the spatial
information (e.g. to generate line maps or velocity fields) rather
than sum the spectra over a large aperture.
For near infrared integral field units where the small field of view
means little or no region of blank sky is included, this could improve
the observing efficiency by nearly a factor of 2.

\section{Conclusions}
\label{sec:conc}

I have presented a method by which the OH emission in the near
infrared bands can be removed.
The technique takes into account variations in the absolute and
relative OH lines as well as instrumental flexure, which may
impact the wavelength scale.
It has been tested using observations performed with the integral
field spectrometer SINFONI.

When the field of view is too small to include blank sky, I have shown
that rather than taking separate blank sky frames at every other
exposure, one can adopt a much more efficient observing strategy.
In principle, one needs to take a sky frame only once per hour in the
H-band where the OH lines are particularly strong; and less often for
K-band observations.
This allows the fraction of time spent on source to be increased from
50\% to as much as 90\% or more.

A version of the code used to perform these procedures is written in
IDL and is available from the author.
In addition, the routines are being implemented in the SINFONI data
processing pipeline by ESO.
The essence of the procedure has also been incorporated into the data
processing software specification \citep{dav06} of KMOS, a near
infrared multi-IFU spectrometer for the VLT \citep{sha05}.

\section*{Acknowledgments}

I thank all those in the Infrared and Submillimetre Group at MPE,
particularly N.~Bouch\'e, who have patiently tested this method.
I am grateful to A.~Modigliani, A.~Davies, M.~Kissler-Patig, and
F.~Eisenhauer for their useful suggestions about the text and
figures. 
I also thank the referee for a number of useful comments and
suggestions.

\label{lastpage}


\begin{thebibliography}{99}

\bibitem[Allington-Smith \& Content(1999)]{all99}
Allington-Smith J., Content R., 1999,
PASP, 110, 1216

\bibitem[Allington-Smith et al.(2006)]{all06}
Allington-Smith J., Content R., Dubbeldam C., Robertson D., Preuss W., 2006
MNRAS, 371, 380

\bibitem[Bonnet et al.(2004)]{bon04}
Bonnet H., et al., 2004,
The ESO Messenger, 117, 17

\bibitem[Davies \& F\"orster Schreiber(2006)]{dav06}
Davies R., F\"orster Schreiber N., 2006, 
ESO/KMOS preliminary design review document VLT-SPE-KMO-146611-001


\bibitem[Eisenhauer et al.(2003a)]{eis03a}
Eisenhauer F., et al., 2003a,
The ESO Messenger, 113, 17

\bibitem[Eisenhauer et al.(2003b)]{eis03b}
Eisenhauer F., et al., 2003b,
in {\em Instrument Design and Performance for Optical/Infrared
  Ground-based Telescopes}, 
eds. Masanori I., Moorwood A., 
Proc. SPIE, 4841, 1548

\bibitem[F\"orster Schreiber et al.(2006)]{for06}
F\"orster Schreiber N., et al., 2006,
ApJ, 645, 1062

\bibitem[Krabbe et al.(2006)]{kra06}
Krabbe A., Larkin J., Iserlohe C., Baraczys M., Quirrenbach A.,
McElwain M., Weiss J., Wright S., 2006,
in {\em Ground-based and Airborne Instrumentation for Astronomy},
eds. McLean I., Masanori I.,
Proc. SPIE, 6296, 151

\bibitem[Larkin et al.(2006)]{lar06}
Larkin J., et al., 2006,
in {\em Ground-based and Airborne Instrumentation for Astronomy},
eds. McLean I., Masanori I.,
Proc. SPIE, 6296, 42

\bibitem[Maihara et al.(1993)]{mai93}
Maihara T., Iwamuro F., Yamashita T., Hall D., Cowie L., Tokunaga A.,
Pickles A., 1993,
PASP, 105, 940

\bibitem[Moorwood et al.(1998)]{moo98}
Moorwood A., et al., 1998,
The ESO Messenger, 94, 7

\bibitem[Oliva \& Origlia(1992)]{oli92}
Oliva E., Origlia L., 1992,
A\&A, 254, 466

\bibitem[Osterbrock et al.(1996)]{ost96}
Osterbrock D., Fulbright J., Martel A., Keane M., Trager S., Basri G.,
1996,
PASP, 108, 277

\bibitem[Rousselot(1997)]{rou97}
Rousselot P., 1997,
``Calculation of the OH infrared spectrum (preliminary report)''

\bibitem[Rousselot et al.(2000)]{rou00}
Rousselot P., Lidman C., Cuby J.-G., Moreels G., Monnet G., 2000,
A\&A, 354, 1134

\bibitem[Sharples et al.(2005)]{sha05}
Sharples R., et al., 2005, 
The ESO Messenger, 122, 2

\bibitem[Wallace(1968)]{wal68}
Wallace L, 1968,
``OH bands in the airglow'', Kitt Peak National Observatory,
Contribution N142

\bibitem[Williams(1996)]{wil96}
Williams P., 1996,
P\&SS, 44, 163

\end{thebibliography}
\end{document}